# Notes on Hybridization in Leaf frogs of the Genus *Agalychnis* (Anura, Hylidae, Phyllomedusinae)

Andrew R. Gray

The Manchester Museum
The University of Manchester
ENGLAND

**Abstract**

Two species of Endangered Leaf-frogs, *Agalychnis moreletii* and *Agalychnis annae*, belonging to the tree frog Subfamily Phyllomedusinae, Genus *Agalychnis*, were hybridized for the first time whilst being maintained in captivity. Previous to this, these allopatric Central American species were considered as being distinctly separate. Crossbreeding following genetic analysis reveals that the two species are extremely closely related, and the hybrid of *A. moreletii* and *A. annae* is presented for the first time. The importance of identifying degrees of genetic variation between species and different populations of the same species, for conservation purposes, is highlighted and discussed.

**Introduction**

In order to successfully conserve a species it is important to know as much as possible about it, particularly its phylogenetic status within the specific taxonomic group involved and its relationship with closely related species. Species referred to in these notes occupy distinct regions of Central America, and until recently their taxonomic relation has remained unclear. However, with the development of molecular technology, and with improvements in captive breeding, there are new opportunities to investigate many aspects of taxonomy that have hitherto remained impossible.

***Agalychnis: Agalychnis annae* and *Agalychnis moreletii***

There are currently 14 recognized species of Leaf frogs belonging to the genus *Agalychnis*: *Agalychnis annae* (Duellman, 1963); *Agalychnis aspera* (Peters, 1873); *Agalychnis buckleyi* (Boulenger, 1882); *Agalychnis callidryas* (Cope, 1862); *Agalychnis dacnicolor* (Cope, 1864); *Agalychnis danieli* (Ruíz-Carranza, Hernandez-Camacho and Rueda-Almonacid, 1988); *Agalychnis granulosa* (Cruz, 1989); *Agalychnis hulli* (Duellman and Mendleson, 1995); *Agalychnis lemur* (Boulenger, 1882); *Agalychnis medinae* (Funkhouser, 1962); *Agalychnis moreletti* (Dumeril, 1853); *Agalychnis psilopygion* (Cannatella, 1980); *Agalychnis saltator* (Taylor, 1955); *Agalychnis spurrelli* (Boulenger, 1913). Of these, 3 species are currently classified as being Endangered or Critically Endangered: *Agalychnis moreletii, Agalychnis annae,* and *Agalychnis lemur*. All species of *Agalychnis* share common characteristics of the genus, and each has distinct morphological features on which it was originally described. These features include the amount of webbing between the toes, distinctive colourations of the hands and feet, flanks, and irises, the presence or absence of reticulated palpebral membranes, and also several osteological differences. Recent revision has led to the inclusion of the Genus *Pachymedusa*, and also *Hylomantis,* now being considered paraphyletic and the synonymy of *Agalychnis* (Faivovich, et al., 2010). *Agalychnis litodryas*, which was originally described from one specimen, is now been considered a synonym of *Agalychnis spurrelii* (Ortega-Andrade, 2008; Faivovich, et al., 2010).

Although the range of certain species, such as *Agalychnis callidryas,* is extensive (Mexico to Panama), the distribution of other species is extremely limited and geographically separate. For example, until recently *A. annae* was considered to be endemic to the highlands of Costa Rica (Savage, 2002). The holotype originated from Tapanti, Cartago Province, Costa Rica, and was originally described by Duellman in 1963 as *Phyllomedusa annae* (Duellman, 1963). Until recently all reported locations for the species were in Costa Rica, but the distribution has now been extended to include the Serrania de Tabasara, west-central Panama (Hertz, et al., 2011).

Another example, with a quite different geographical distribution to *A. annae*, is *Agalychnis moreletii,* a species found in humid montane tropical forests ranging from Southern Mexico to central Guatemala, El Salvador, north-western Honduras, and Belize (Santos-Barrera et al., 2004). This *Agalychnis*, which appears to be the ecological equivalent to *A. annae* in Northern Central America (Duellman, 1970), differs in its colouration from that species: *A. annae* has yellow-orange irises and blue flank colouration; *A. moreletii* has very dark red irises and no blue flank colouration. The difference in colouration was used as the main distinctive character for separating the two species in 1963 (Duellman, 1963). Prior to this, specimens collected in Costa Rica were considered to be *Agalychnis (Phyllomedusa) moreletii* (Taylor, 1952).

Morphologically, and from an osteological perspective, the two species are extremely similar. The skulls of both *A. annae* and *A. moreletii* are barely distinguishable when several specimens of each species of similar age and size are compared. Both have skulls that are considerably deeper than any other *Agalychnis* species and also show the greatest amount of ossification of the sphenoid (Duellman, 1970). Here it is interesting to note that, prior to any molecular analysis, a cladistic phylogenetic analysis of the Genus *Agalychnis* based on a matrix containing 12 phenotypic characters revealed five equally parsimonious trees. Out of these, only one clear polytomy emerged: *A. moreletii* and *A. annae* (Duellman, 2001). With the inclusion of other species in the genus due to recent revisions (and excluding the former *A. calcarifer* and *A. craspedopus*, which are both now in the genus *Cruziohyla*), many of the phenotypic characters have been redefined (Faivovich et al., 2010). However, apart from differences in coloration, the two species concerned continue to share the same defining characteristics. When a comprehensive matrix using different characters associated with phyllomedusine breeding biology was carried out (Faivovich, *et. al.*, 2010), *A. annae* and *A. moreletii* shared all reproductive characteristics, which in turn separated them from all other *Agalychnis* species. The characters were larval development in ponds, egg-less capsules absent, leaf-folding behavior present, oviposition on leaves, epiphytes, and roots. There is little discernable difference in egg clutch size (Gomez-Mestra, *et al.*, 2008), although this can vary considerably between *Agalychnis* species. Further, the tadpoles of the two species appear almost indistinguishable, particularly when compared at the same stage of development (personal observation), but they differ noticeably from all other *Agalychnis* tadpoles by having a shorter tail, deeper body, and a protruding snout (Duellman, 1970).

In the past few years the Hylid subfamily Phyllomedusinae has been the focus of much phylogenetic research and recent studies based on molecular analysis have proved extremely valuable in clarifying the taxonomic status of many species (Kerfoot, 2003, Faivovich et al., 2005, Wiens et al., 2005, Moen and Wiens, 2008, Faivovich, et al., 2010). A study carried out by Crook (2007) at Manchester University as part of her Masters Degree used phylogenetic analysis to compare mitochondrial *16s* and cytochrome b (*cytb*) gene sequences from different Phyllomedusine species. This confirmed Kerfoot's original results (Kerfoot, 2003). The work of Crook also supports the hypothesis of Faivovich, *et. al.* (2010) and the recent re-classification that places the species previously known as *Pachymedusa dacnicolor* in the Genus *Agalychnis* (Faivovich, et al, 2010). In 2003, Kerfoot confirmed the species now known as *Cruziohyla calcarifer* as being distinctly separate from *Agalychnis* and *Phyllomedusa* (Kerfoot, 2003). Assessment of the genetic relationship between the species *A. annae* and *A. moreletii* was unavailable to him at the time. More recently, however, detailed phylogenetic studies that have included both *A. annae* and *A. moreletii* have shown that they are more closely related to each other than to any other species in the genus (Gomez-Mestre et al., 2008; Faivovich, et al., 2010). Further, the study by Crook (2007) indicated that *A. annae* and *A. moreletii* provided branch lengths separating the species of just 0.047 and 0.029 (Crook, 2007: Figure 1), comparable to the variation between distinct populations of *Agalychnis lemur* from Costa Rica and Panama, which are currently classified as the same species.

Following this genetic assessment, and to investigate genetic compatibility, a crossbreeding experiment between the two species was carried out at Manchester Museum in 2010 in which unassisted hybridization between specimens of and *A. moreletii* from Guatemala (Plate 1) and *A. annae* from Costa Rica (Plate 2) was achieved.

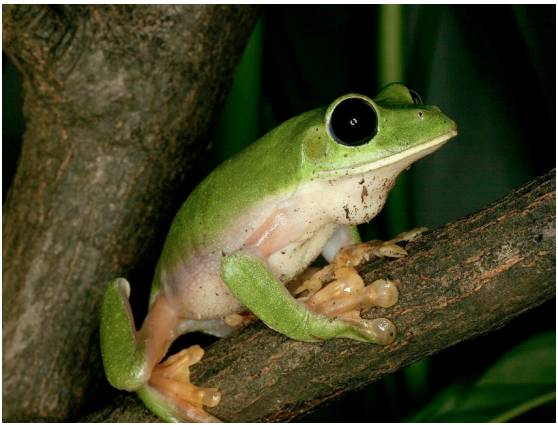

**Plate 1.** *Agalychnis moreletii* (Guatemalan specimen) © Tobias Eisenburg

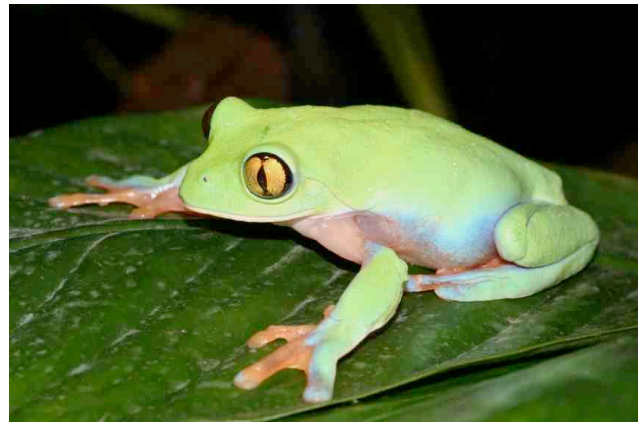

**Plate 2.** *Agalychnis annae* (Costa Rican specimen) © Andrew R. Gray

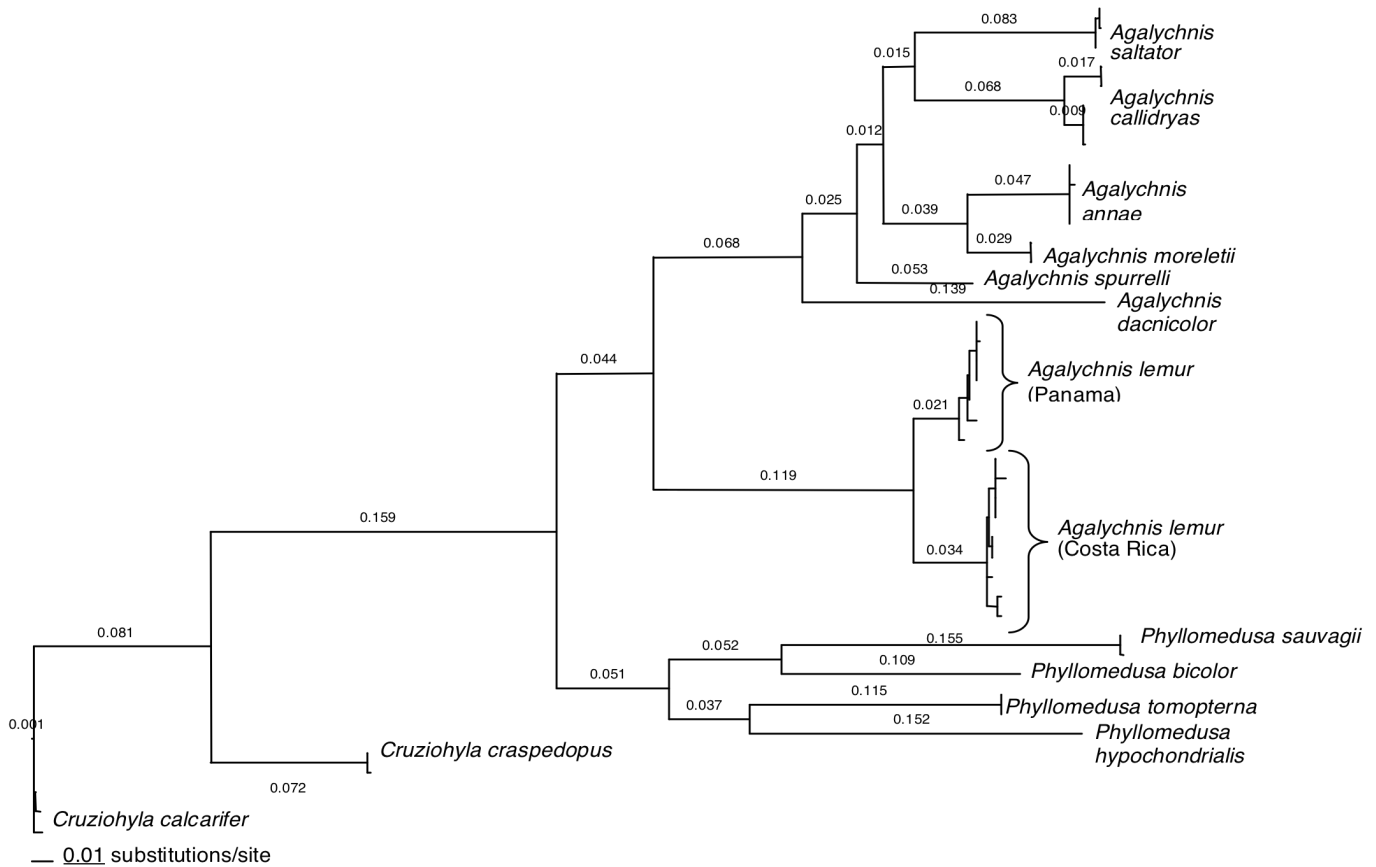

**Figure 1.** Bayesian concensus phylogram with *Cruziohyla, Phyllomedusa* and *Agalychnis*. Phylogram is rooted with the outgroup taxon *Cruziohyla calcarifer*. Values represent the number of substitutions for site for that lineage (Redrawn from Crook, 2007).

### The Hybrid:

Structurally, and from an osteological perspective, the hybrid of *A. annae* and *A. moreletii* is barely distinguishable from either parental species. This is to be expected, as they themselves are structurally extremely similar. The parent species share characteristics that separate them from other species of the genus *Agalychnis*: a very small bone called the squamosal is in narrow contact with the crista parotica; the cloacal sheath is long and directed ventrally; the iris is red or yellow/orange. The hybrid of *A. annae* and *A. moreletii* is no different in this respect. The hybrid exhibits colouration characteristics of both species: the iris colouration is very dark red, most like *A. moreletii*, and the flanks and thighs have purple to blue coloration, most like *A. annae*. However, unlike the extensive blue colouration of the flank markings seen in *A. annae*, the purplish-blue in the hybrid forms a narrow band between the green dorsal coloration and the white underside (Plate 3). A purplish hue also extends along the external edging of the arms and legs. The hands and feet are orange, and this is more intense than was seen in either parent. At the time of writing, the hybrid is being maintained in captivity, and the final adult size is so far unknown.

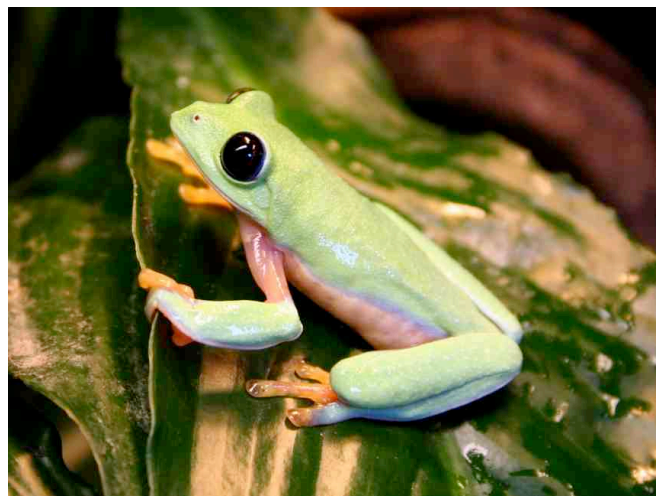

**Plate 3.** A hybrid of *Agalychnis moreletii/Agalychnis annae* (D1285). © Andrew R. Gray, The Manchester Museum.

**Discussion**

Hybridization experiments with amphibians have proved particularly useful in providing evidence for the inheritance of genes, including certain colour pattern traits (Duellman & Trueb, 1986). The cross mating of *A. annae* and *A moreletii* produced a hybrid with purplish-blue flank and thigh markings, a feature normally associated with *A. annae* (Kubicki, 2004). The hybrid also has a very dark red iris, a feature of *A. moreletti* (McCranie and Wilson, 2002). The species concerned in this study are allopatric, but natural crossbreeding in closely related sympatric species belonging to the subfamily Phyllomedusinae has shown that the hybrids generally do have intermediate colour patterns (Haddad, et al., 1994).

At the time of writing, the hybrid is not fully grown, but it is expected to attain a size similar to its parents; females of both species reach a similar size, although there seems to be some differences in the extent of sexual dimorphism. The original 40 male specimens of *A. annae*, collected from Cartago Province, Costa Rica, by Edward Taylor on 26th/27th August 1947 as *A. moreletii*, are now held in different museums (eg. 26 specimens are held in The Field Museum and 15 specimens in Kansas University Museum). In his work relating to the specimens collected, Taylor stated that the males attained 54-55mm SVL, and seem smaller that Mexican *A. moreletii* (Taylor, 1952). Larger male specimens and females of *A. annae* have since been found, including those also collected from Cartago as types in 1961. In the description of *A. annae* by Duellman (1963), the written sizes of 35 male *A. annae* specimens are given as being 55.9 - 65.7mm SVL, along with 5 female specimens having snout-vent lengths of 81.6 - 84.2mm (Holotype and paratypes of *A. annae* at Kansas University total 40 (KU 64020-64060). However, there appears to have been some confusion in Duellman's paper concerning sizes and the comparison of *A. annae* with *A. moreletii*: the sizes given in the text for the 35 male *A. annae* specimens appear in the comparison table under the 25 male *A. moreletii*, and sizes reaching 73.9mm appear for the *A. annae* males. It has since been confirmed that male *A. annae* specimens do reach such a large size: 75mm in SVL and females 85mm SVL (Arguedas, 2010). To date, male *A. moreletii* are known to reach 66 mm SVL (Briggs, 2010) and females 85 mm SVL (Lee, 1996).

Hylid species are recognised to have great differences in size, iris colouration, and flank colouration and pattern (Duellman, 1970). Within *Agalychnis*, there is a great amount of variation in flank markings in *A. callidryas*, a species that occurs from Mexico to Panama. Depending on the geographical locality of the population, the colouration of the flanks, thighs, and concealed surfaces in this species can vary from blue to orange (Robertson and Zamudio, 2009). The colouration of the flanks, thighs, and concealed surfaces (including hands and feet) from different *A. moreletii* populations also varies geographically; little or no orange colouration in specimens from areas in Guatemala (Plate 1) and areas of neighboring El Salvador (Plate 4), orange in specimens from areas of Mexico (Lee, 1996), vivid orange in specimens from areas of Belize (Briggs, 2010: Plate 5). In *A. annae* from Costa Rica the flanks and posterior part of the thighs are purplish-blue (Kubicki, 2004) (Plate 2). Although little is currently known about the *A. annae* population occurring in west-central Panama, the initial specimen found has purplish-blue colouration to the flanks and thighs. However, the specimen also has a covering of distinctly raised white-coloured pustules on the dorsal surface, most similar in structure to those seen in some specimens of *Agalychnis dacnicolor* from Mexico (personal observation of *A. dacniolor* and assessment of photograph of Panamanian *A. annae* taken by A. Hertz: see Hertz, et al., 2011). *A. annae* specimens from Costa Rica do not have these white pustular spots, and are characterized by having a smooth dorsal surface (Savage, 2002), which is pale green and shows little variation (Duellman, 1963). Interestingly, having a scattering of white flecks on the dorsum is a distinctive feature of *A. moreletii* specimens from some populations, including those from Honduras (Duellman, 1970).

In some other groups of neotropical frogs, such as those belonging to the Dendrobatidae Genus *Oophaga*, the most extreme polymorphism in patterning and colour can be seen. In some species, such as *Oophaga pumilio*, specimens from different geographically separated localities range in colour from bright red, orange, yellow, to green (Daly and Myers, 1967). In this species, which occurs from Nicaragua to Panama, some specimens from Costa Rica have solid blue markings on their legs (Savage, 2002). In the population of frogs from Tierra Ocsura, Panama, almost completely blue specimens occur (Hagemann and Prohl, 2007). The blue skin of the *O. pumilio* lacks the xanthophore layer (Frost-Mason, et al, 1972). The same skin pigment layer loss is responsible for the blue colouration seen in *A. annae.*

Several factors are considered to be responsible for phenotypic variation in amphibians, including natural and sexual selection, geographical isolation, and genetic drift (Wang and Shaffer, 2008). However, until recently, phenotypic variation is an aspect of amphibian evolutionary biology that has remained extremely understudied. Several hypotheses regarding causes of phenotypic variation have recently been tested, and in the brightly coloured diurnal dart frog species *O. pumilio*, the hypothesis of colour pattern divergence being due to isolation by distance (Wright, 1943) has been rejected in favour of one supporting divergent selection, and reproductive isolation being responsible for producing differently coloured populations (Wang & Summers, 2010). Although female choice in dart frogs may be based on aposematic colouration (Maan & Cummings, 2008), other recent studies focusing on more cryptic and nocturnally active Central American neotropical frog species, such as *Physalaemus pustulosus*, show both reproductive isolation through distance and time of divergence between populations as factors accounting for differences in the genetic structure of populations (Pröhl, et al., 2010). However, it should be emphasised that reproductive behaviour of many neotropical frogs is very different than that of dendrobatids, where territoriality of males and female mate choice may play a more important role in their reproduction (Stebbins and Cohen, 1995).

Agalychnis frogs are known for their cryptic colouration when at rest, and extremely bright colouration when active (Gray and Drury, 2004). Such brightly coloured flash marking in phyllomedusines are used to avoid predation (Duellman and Trueb, 1986). However, males are known to use their flank and leg colourations when visual signaling, as when body posturing and leg-waving in conspecific territorial communication (Gray, 2002). Recent studies have shown that, rather than genetic drift, both localized selection and geographic barriers contribute to colour divergence among *Agalychnis* populations (Robertson and Zamudio, 2009). Duellman suggests that *A. annae* and *A. moreletti* probably evolved from their ancestral stock as a result of geographical isolation, correlated with the elevation of the Talamanca mountain range in lower Central America and the highlands of nuclear Central America. (Duellman, 2001).

Within *Agalchnis*, several species stand out as having evolved to inhabit high mountains: *Agalychnis moreletti* (450-2000m); *Agalychnis annae* (780-1,650m); *Agalychnis lemur* (440-1,600m) (Duellman, 2001; Savage, 2002). Within the genus these are the species that are considered to be the most at risk of becoming extinct, as reflected in their current conservation status: *A. moreletii* and *A. lemur* are both classified as Critically Endangered and *A. annae* as Endangered on the IUCN Red Data List. Over the past 30 years, populations of all three species have completely disappeared from areas where they were once common, particularly in the highlands of Central America where cooler temperatures and Chytridiomycosis caused by the fungus *Batrachochytrium dendrobatidis* both appear to have played their part in their demise (Pounds et al., 2006; Lips, 2006). Surveys have indicated that *A. moreletii* has now completely disappeared from several areas where it was once abundant, particularly in Honduras, and also Mexico (Lips, et al 2004). What remains of the Costa Rican population of *A. annae* now only survives in small pockets at the lower altitudes of its former range, in the Central Valley and around the Capital, San Jose (Pounds, et al., 2008). The same is true for *A. lemur*, where the Costa Rican populations are now known from just 2 or 3 sites.

It is likely that specific environmental conditions in the areas where remaining populations survive contribute to the species' current preservation, and that local temperature and humidity levels, driven by precipitation and sunlight, are important in their resistance to the chytrid. As Robertson and Zamudio (2009) proposed, dispersal barriers and geographical separation have probably brought about the genetic and phenotypic divergence in Agalychnis frogs (Robertson and Zamudio, 2009). Presumably, these same barriers are also partly responsible for protecting individual populations from environmentally associated threats. It follows that individual populations that have had the opportunity to adapt to favourable conditions may be at less risk than others, heightening the importance of identifying differences in populations in relation to conservation efforts. There is also real concern that certain populations may disappear before their distinctiveness has even been established.

A measurement to complement the species concept was developed some time ago in an effort to assist such conservation, the "evolutionary significant unit" (ESU). An ESU is 'A defined population that is considered morphologically and genetically different from related ESU's, caused by either past restriction of gene flow, locally adapted phenotypic traits, or current geographical separation' (Connor and Hartl, 2004). If populations are considered to be ESU's, it is generally agreed that they should be treated separately for the purposes of conservation (Ryder, 1986).

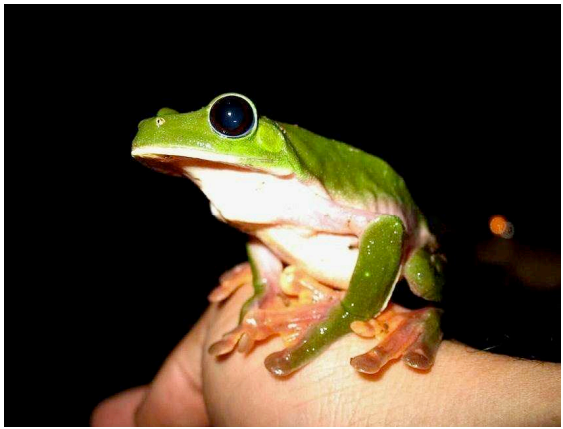

**Plate 4.** *Agalychnis moreletii*, El Imposible National Park, Western El Salvador. © Tyler Lawson; Lawson et al., (2011).

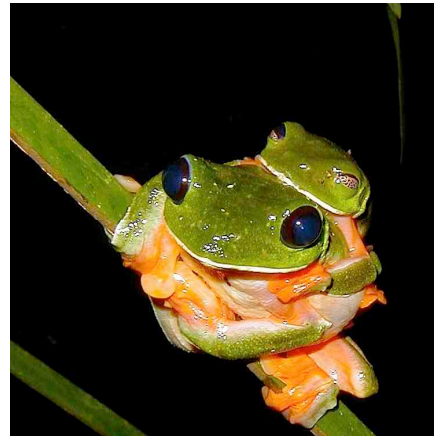

**Plate 5.** *Agalychnis moreletii*, Las Cuavas, Chiquibul Forest Reserve, Belize. © Venetia Briggs; Briggs (2010).

## Conclusion

Crossbreeding has shown that the non-sympatric species *Agalychnis annae* and *Agalychnis moreletti* are extremely closely related. However, they should continue to be considered as separate, both as allopatric taxa and for conservation purposes. It is also important to recognize the levels of similarity observed in distinct populations of other closely related species, such as *Agalychnis lemur* from Costa Rica and Panama: if conservation is of main importance, it follows that the separate populations of *A. lemur* should also be conserved for the future as distinct entities.

Future studies should also focus on assessing the levels of variation between different populations of the same species. Extensive research has already begun in this area, such as the detailed work by Pröhl, *et al.*, (2010) with the neotropical frog *Physalaemus pustulosus,* and by Robertson and Zamudio (2009) with *A. callidryas*. Focusing on an endangered species would add real conservation value. Suggested future research relating to some of the issues highlighted in this work might include determining levels of geographic variation and genetic diversification in *A. moreletii* from countries where they show significant contrast in colour, comparing the newly discovered population of *A. annae* in Panama with those from Costa Rica, and investigation of the individuality of *A. lemur* populations. Whatever action is felt most appropriate, more importance should be placed on sampling specimens from throughout the species' distribution to assist further status and conservation assessments. As phylogenetic assessment opportunities develop, agreement from within the scientific community on standardization of the level of genetic divergence required to define an amphibian 'ESU' would be highly beneficial to taxonomists and conservation biologists alike.

Although crossbreeding experiments can provide much useful information, the author recognizes the ethical considerations involved. This work was carried out under controlled conditions, and as a one-off for the benefit of the scientific discipline. It is hoped that these notes will provide useful information to others concerned with furthering our knowledge of amphibian phylogenetics and those working to conserve all the endangered species discussed.

## Acknowledgements

I would like to thank the following people for supporting my interest in working with Phyllomedusine species over the years and in facilitating this work: Javier Guavara Sequiera, Federico Bolaños, Miguel Solano, Brian Kubicki, Ron Gagliardo, Karl-Heinz Jungfer, Joseph Bagnara, Morley Reid, Luis Coloma, Jose Hernández, Chris Kerfoot, Tara Crook, Cathy Walton, Alan Pounds, Heinz Hoffman, Darren Smy, Adam Bland, Tobias Eisenberg, Andreas Hertz, Gunther Köhler, Tyler Lawson, Venetia Briggs, Viviana Arguedas, Kevin Healey, Laurence Cook.